\documentclass[fleqn,usenatbib]{mnras}

\usepackage{graphicx}
\usepackage[export]{adjustbox}
\usepackage{times}
\usepackage{grffile} %
\usepackage{newtxtext,newtxmath}
\usepackage{bm}
\usepackage{physics}
\usepackage{cleveref}
\usepackage{xspace}
\usepackage{booktabs}
\usepackage[flushleft]{threeparttable}
\usepackage{siunitx}
\usepackage[super]{nth}
\DeclareSIUnit{\hred}{\mathit{h}}
\DeclareSIUnit{\Msun}{M_\odot}
\DeclareSIUnit\pc{pc}
\DeclareSIUnit\kpc{kpc}
\DeclareSIUnit\Mpc{Mpc}
\DeclareSIUnit\lightyear{ly}
\usepackage[table]{xcolor}
\usepackage{soul}
\usepackage[normalem]{ulem}
\usepackage{enumerate}
\usepackage{tabularx}
\usepackage{multirow}
\usepackage{rotating}
\usepackage{caption}
\usepackage{subcaption}

\definecolor{bubbles}{rgb}{0.91, 1.0, 1.0}
\definecolor{aquamarine}{rgb}{0.5, 1.0, 0.83}
\definecolor{bubblegum}{rgb}{0.99, 0.76, 0.8}
\definecolor{bluebell}{rgb}{0.74, 0.74, 0.92}
\definecolor{dollarbill}{rgb}{0.72, 0.93, 0.6}

\newcommand{\Msun}{\ensuremath{\mathrm{M}_\odot}}

\crefname{table}{Table}{Tables}
\Crefname{table}{Table}{Tables}

\crefname{equation}{Eq.}{Eqs.}
\Crefname{equation}{Equation}{Equations}

\crefname{figure}{Fig.}{Figs.}
\Crefname{figure}{Figure}{Figures}

\crefname{section}{Section}{Sections}
\Crefname{section}{Section}{Sections}
\crefname{appendix}{Appendix}{Appendices}
\Crefname{appendix}{Appendix}{Appendices}

\newcommand{\ramses}{\textsc{ramses}\xspace}

\newcommand{\genetic}{\textsc{genetIC}\xspace}
\newcommand{\adaptahop}{\textsc{adaptahop}\xspace}

\newcommand{\AM}{angular momentum\xspace}

\renewcommand{\vec}[1]{\bm{#1}}

\title[Controlling stellar angular momentum]{%
    Stellar angular momentum can be controlled from cosmological initial conditions
}

\author[C.~Cadiou et al.]{Corentin~Cadiou$^{1}$\thanks{c.cadiou@ucl.ac.uk},
Andrew~Pontzen$^{1}$,
Hiranya~V.~Peiris$^{1,2}$
\\
$^{1}$Department of Physics and Astronomy, University College London, Gower Street, London WC1E 6BT, United-Kingdom\\
$^{2}$The Oskar Klein Centre for Cosmoparticle Physics, Department of Physics, Stockholm University, AlbaNova, Stockholm SE-106 91, Sweden\\
}

\date{Accepted XXX. Received YYY; in original form ZZZ.}
\pubyear{2022}

\begin{document}%
\label{firstpage}%
\pagerange{\pageref{firstpage}--\pageref{lastpage}}
\maketitle

\begin{abstract}
    The angular momentum of galaxies controls the kinematics of their stars, which in turn drives observable quantities such as the apparent radius, the bulge fraction, and the alignment with other nearby structures.
    To show how angular momentum of galaxies is determined, we build high (\SI{35}{pc}) resolution numerical experiments in which we increase or decrease the angular momentum of the Lagrangian patches in the early universe.
    We simulate three galaxies over their histories from $z=200$ to $z=2$, each with five different choices for the angular momentum (fifteen simulations in total).
    Our results show that altering early-universe angular momentum changes the timing and orbital parameters of mergers, which in turn changes the total stellar angular momentum within a galaxy's virial radius in a predictable manner. Of our three galaxies, one has no large satellite at $z=2$; in this case, the specific angular momentum is concentrated in the central galaxy.
    We modify its stellar angular momentum over \SI{0.7}{dex} (from \num{61} to \SI{320}{kpc.km.s^{-1}}) and show that this causes its effective radius to grow by $\SI{40}{\percent}$, its $v/\sigma$ parameter to grow by a factor $\times 2.6$ and its bulge fraction to decrease from \num{0.72} to \num{0.57}.
    The ability to control angular momentum will allow future studies to probe the causal origin of scaling relations between galaxy mass, angular momentum and morphology, and to better understand the origin of galactic intrinsic alignments.
\end{abstract}


\section{Introduction}%
\label{sec:introduction}

Angular momentum plays a major role in galaxy formation. In spiral galaxies, it dictates the size and alignment of the disk; elliptical galaxies, by contrast, are dispersion-supported and have eight times lower stellar specific angular momentum $j_\star$, explaining the morphological distinction \citep{1983IAUS..100..391F,romanowsky_ANGULARMOMENTUMGALAXY_2012, fall_AngularMomentumGalaxy_2018,harrison_KMOSRedshiftOne_2017,espejosalcedo_MultiresolutionAngularMomentum_2022}.
Moreover, the angular momentum of neighbouring galaxies is partially correlated.
This effect, known as intrinsic alignment \citep{troxel_IntrinsicAlignmentGalaxies_2015}, needs to be properly modelled to disentangle it from cosmic shear and thus ensure the success of upcoming cosmological weak-lensing surveys (\emph{Euclid},~\citealt{laureijs_EuclidDefinitionStudy_2011}; Vera Rubin Observatory,~\citealt{ivezic_LSSTScienceDrivers_2019}).

Early explanations for the origin of galactic angular momentum assumed that the gas and dark matter (DM) within a given halo acquire identical spin, driven by tidal torques from the large-scale environment \citep{peebles_OriginAngularMomentum_1969,doroshkevich_SpaceStructurePerturbations_1970,white_AngularMomentumGrowth_1984}.
Even though gas radiatively cools to form stars, in doing so it may conserve its angular momentum, so that stars in a galaxy would inherit the DM halo's spin \citep{fall_FormationRotationDisc_1980,mo_FormationGalacticDiscs_1998}. Recent work shows that, at the population level, the specific angular momentum distributions of simulated galaxies and DM halos are indeed strikingly similar \citep{danovich_FourPhasesAngularmomentum_2015}.
However, on a per-object basis, their magnitudes are poorly correlated \citep{jiang_DarkmatterHaloSpin_2019} and they show significant misalignments, with the extent of the mismatch depending on redshift, mass, or the central/satellite nature of the host \citep{tenneti_GalaxyShapesIntrinsic_2014,velliscig_AlignmentShapeDark_2015,chisari_GalaxyhaloAlignmentsHorizonAGN_2017}.

Multiple explanations have been put forward to explain how the galaxy spin may decouple from the host halo spin.
First, gas can cool even outside the halo and therefore the dark matter and baryons accreted into a halo do not necessarily originate in the precise same patch of the early universe \citep{pichon_RiggingDarkHalos_2011,kimm_AngularMomentumBaryons_2011,liao_SegregationBaryonsDark_2017}. This in turn implies that the tidal torques differ between the two components.
At high redshift, and for sufficiently low-mass galaxies, most of the gas and angular momentum flows into galaxies through cold flows  \citep{dekel_GalaxyBimodalityDue_2006,tillson_AngularMomentumTransfer_2015,stewart_HighAngularMomentum_2017}, allowing the gas to acquire a larger angular momentum than DM prior to accretion \citep{stewart_ObservingEndCold_2011,danovich_FourPhasesAngularmomentum_2015,cadiou_GravitationalTorquesDominate_2021}.

A second source of galactic angular momentum is from mergers.
The orbital angular momentum from mergers causes the magnitude and orientation of the DM halo spin to change near-instantaneously \citep{vitvitska_OriginAngularMomentum_2002,benson_RandomwalkModelDark_2020a}, but the effects on the galaxy are slower and more complex. The orientation and magnitude of spin in the post-merger disk eventually depends on the fraction of orbital angular momentum conserved during the merger, and on the efficiency and angular momentum richness of the post-merger star formation that rebuilds the disk. Consequently, the final result of a galaxy merger can depend on the gas fraction and morphology of the progenitors \citep{barnes_TransformationsGalaxiesII_1996,naab_StatisticalPropertiesCollisionless_2003,lotz_GalaxyMergerMorphologies_2008,athanassoula_FormingDiskGalaxies_2016,governato_FormingLargeDisc_2009,garrison-kimmel_OriginDiverseMorphologies_2018},
 the orbital configuration of the merging system \citep{martin_RoleMergersDriving_2018,jackson_WhyExtremelyMassive_2020,zeng_FormationMassiveDisc_2021},
and feedback and cooling processes at play \citep{robertson_MergerdrivenScenarioCosmological_2006,cox_KinematicStructureMerger_2006}.

Even once gas is inside a halo, a variety of processes determine whether it is incorporated into the disk, and therefore how it changes the stellar angular momentum. Cold flows may be hydrodynamically disrupted \citep[see e.g.][]{mandelker_InstabilitySupersonicCold_2016,cornuault_AreCosmologicalGas_2018,padnos_InstabilitySupersonicCold_2018,mandelker_InstabilitySupersonicCold_2019,aung_KelvinHelmholtzInstabilitySelfgravitating_2019,mandelker_InstabilitySupersonicCold_2020} or be blown away by feedback \citep{dubois_BlowingColdFlows_2013,ramsoy_RiversGasUnveiling_2021}. Centrally-concentrated supernova feedback may preferentially expel low-angular momentum gas from the centre of galaxies, resulting in an amplification of the spin of the galactic disk \citep{brook_HierarchicalFormationBulgeless_2011,ubler_WhyStellarFeedback_2014}. Finally, accretion of counter-rotating material may lead to a rapid contraction of a galactic disk \citep{dekel_WetDiscContraction_2014,zolotov_CompactionQuenchingHighz_2015}, generating a starburst and associated feedback energy that can expel remaining gas.

In summary, the final angular momentum of galaxies depends on \emph{(a)} the angular momentum originating in the cosmological initial conditions, \emph{(b)} how this angular momentum is transported from cosmological to galactic scales \emph{via} smooth accretion and mergers, and \emph{(c)} what fraction of this angular momentum remains in the galaxy and its disk.
These dependencies are complex, raising the question of whether or not galaxies retain a clear memory of angular momentum generated by tidal torques.
Beyond its inherent interest in galaxy formation theory, answering that question is ultimately essential for understanding the amplitude of intrinsic alignments for weak lensing.

In this paper, we test whether galaxies retain a memory of their cosmologically-acquired angular momentum.
We perform a numerical experiment in which we resimulate three galaxies five times each, systematically modifying the angular momentum acquired by the baryons in the linear early universe. This is accomplished by creating four ``genetically modified'' versions of the initial conditions, using the technique of \cite{roth_GeneticallyModifiedHaloes_2016} to ensure the changes are minimal and consistent with the cosmological statistics. The approach was recently extended to the case of angular momentum by \cite{cadiou_AngularMomentumEvolution_2021}. After resimulating the additional initial conditions, 
we measure the stellar angular momentum in the galaxy at $z\approx 2$, and assess the degree to which it correlates with the angular momentum in the initial conditions. 
By performing this experiment in a zoom cosmological simulation, we capture angular momentum acquisition at cosmological scale, its transport into galaxies and the relevant physical processes in play at galactic scale.
We employ cosmological zoom-in simulations with state-of-the-art physics using the code \ramses{} \citep{teyssier_CosmologicalHydrodynamicsAdaptive_2002}.

The structure of the paper is as follows.
In \cref{sec:method}, we describe our numerical setup and how we genetically modify the initial conditions to alter the angular momentum of three simulated galaxies.
We then present our results in \cref{sec:results}, before discussing their implications  in \cref{sec:discussion}.

\section{Method}%
\label{sec:method}

\subsection{Simulations}%
\label{sec:simulations}

 All our simulations are performed within a cosmology that has a total matter density of $\Omega_\mathrm{m} = 0.3089$, a dark energy density of $\Omega_\Lambda = 0.6911$, a baryonic mass density of $\Omega_\mathrm{b}= 0.0486$, a Hubble constant of $H_0 = \SI{67.74}{\km.s^{-1}.Mpc^{-1}}$, a linear variance at \SI{8}{Mpc} $\sigma_8 = 0.8159$, and a power spectrum index of $n_s = 0.9667$, compatible with a {\it Planck} 2015 cosmology~\citep{planckcollaboration_Planck2015Results_2015}.

 We first generate a $512^3$ dark matter-only simulation of a cosmological box of comoving side length \SI{100}{\hred^{-1}.Mpc}, using initial conditions generated by \genetic{} \citep{stopyra_GenetICNewInitial_2020}, which enables the construction of accurate genetic modifications. From this, we select three halos of virial mass $M_\mathrm{vir} \approx \SI{e12}{\Msun}$ at $z=2$ for re-simulation at high resolution.
 Galaxies of this mass range and this redshift display a diverse range of morphologies \citep[see e.g.][]{tamburri_PopulationEarlytypeGalaxies_2014}, which allows us to study the onset of the Hubble sequence.
 We select three halos, hereafter halos A, B and C, with no major merger since $z=2.5$ and no massive nearby neighbour within \SI{500}{kpc} in the low-resolution DM-only simulation.
 The halos are chosen `blind', i.e.\ randomly from halos of the appropriate mass range matching the merger and neighbour criterions.
 When increasing the resolution, mergers below the resolution limit of the low-resolution may appear, as we discuss below.
 In each of the three cases, we identify the Lagrangian patch of particles extending out to three virial radii, and populate the corresponding regions of the initial conditions with high-resolution DM particles.

We perform these three simulations with hydrodynamics using the adaptive mesh refinement code \textsc{ramses}~\citep{teyssier_CosmologicalHydrodynamicsAdaptive_2002}, adopting a minimum cell size and gravitational softening of \SI{35}{pc}, and a mass resolution of $M_\mathrm{DM}=\SI{1.6e6}{\Msun}$ and $M_\star = \SI{1.1e4}{\Msun}$ for DM and stars respectively.
We employ Monte-Carlo tracer particles to recover the Lagrangian history of the baryons, as described in~\cite{cadiou_AccurateTracerParticles_2019}. We use sub-grid baryonic physics following the approach of NewHorizon~\citep{dubois_IntroducingNEWHORIZONSimulation_2021}. In brief, star formation is allowed above a gas density of $n_0=\SI{10}{cm^{-3}}$ with a Schmidt law; the stellar population is sampled with a~\cite{kroupa_VariationInitialMass_2001} initial mass function; and the mass loss fraction from supernovae explosions is $\eta_{\rm SN}=\SI{32}{\percent}$ with a metal yield (mass ratio of the newly formed metals over the total ejecta) of $0.05$. %
Type II supernovae are modelled with the mechanical feedback model of~\cite{kimm_SimulatingStarFormation_2015} with a boost in momentum due to early UV pre-heating of the gas following~\citep{geen_DetailedStudyFeedback_2015}. The simulations also track the formation of supermassive black holes and their energy release through AGN feedback.
Further details about the simulations can be found in \cref{sec:technical_details}.
\begin{figure*}
    \includegraphics[width=\textwidth]{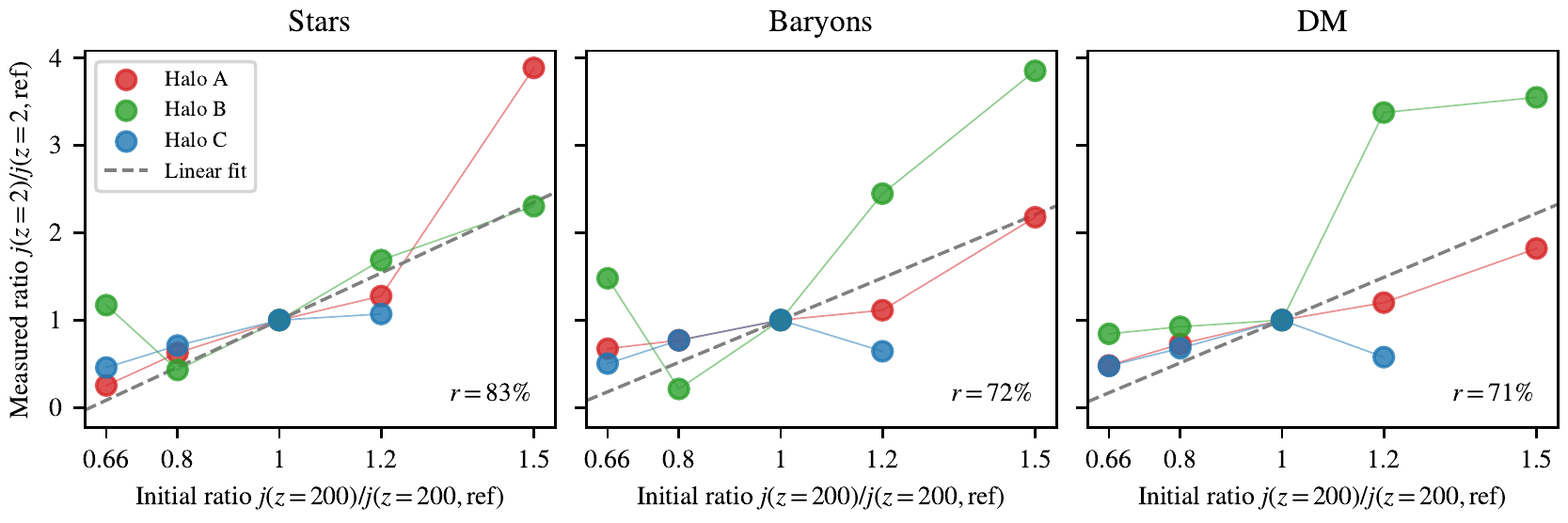}
    \caption{
        We measure the stellar (\emph{left panel}), baryonic (\emph{central panel}) and DM (\emph{right panel}) angular momentum within $R_\mathrm{vir}$ at $z=2$ and compare it to the initial angular momentum change of the baryonic Lagrangian patch at $z=200$.
        The Pearson correlation coefficient between the two quantities is reported in each panel.
        We can control the final stellar angular momentum within $R_\mathrm{vir}$ through modifications in the initial conditions; the baryons and DM angular momentum also respond in a correlated way.
        In the case of galaxies A$\times 1.5$ and B$\times 0.66$, the stellar angular momentum does not change linearly. Galaxy A$\times 1.5$ is undergoing a 1:2 merger, with the companion reaching its first apocenter at $z=2$. Galaxy B$\times 0.66$ has a complex series of mergers, as described in the text.
    }\label{fig:AM_change__vs__stars_AM}
\end{figure*}

We extract galaxy and halo catalogues using \adaptahop{} \citep{aubert_OriginImplicationsDark_2004}, using the `Most massive Sub-node Method' and the parameters proposed in~\cite{tweed_BuildingMergerTrees_2009}. The density is computed from the 20 nearest neighbours and we use a linking length parameter of $b=0.2$.
For each galaxy, we compute the half-mass radius $R_{1/2}$, defined as the radius that contains half the total stellar mass, $M_\star(<R_{1/2}) = M_\star(<R_\mathrm{vir})/2$.
We then perform a kinematic decomposition to compute the bulge-to-total mass ratio, $B/T$.
To that end, we first compute the stellar angular momentum and project all particles in cylindrical coordinates, where the $z$ axis is parallel to the angular momentum vector.
We include the mass of particles in the kinematic bulge if the magnitude of their tangential velocity satisfies $v_\theta^2 < v_r^2 + v_z^2$. 
Finally, we also compute the velocity dispersion parameter, $v/\sigma \equiv \langle v_\theta \rangle / \sqrt{\langle v^2 \rangle}$, where the average is a mass-weighted sum over all stellar particles. 

\subsection{Angular momentum modifications}%
\label{sec:AMGM}

With three galaxy simulations in hand, we next resimulate each with a variety of different angular momenta. We apply the genetic modification approach \citep{roth_GeneticallyModifiedHaloes_2016}, extended for the case of angular momentum as described by \cite{cadiou_AngularMomentumEvolution_2021}. As per \cite{roth_GeneticallyModifiedHaloes_2016}, the appropriate region in the early universe is systematically modified while maintaining consistency with the Gaussianity and power spectrum specified by cosmology. The changes made to the field are minimal so that other structures and large scale filaments are nearly unaffected by the modifications to any given galaxy.  Changes that alter the angular momentum consist of distortions to the local tidal fields. For example, an elongated region of the initial conditions can be `spun up' by adding a small overdensity on its leading edge and corresponding underdensity on the trailing side. For a formal derivation and presentation of the method, see  \cite{cadiou_AngularMomentumEvolution_2021}.

To enable the modification, first a suitable patch of the early universe must be identified in which the angular momentum will be changed. %
The reference galaxies have a dark matter spin parameter \citep{bullock_UniversalAngularMomentum_2001a} $\lambda \equiv j/\sqrt{2}V_\mathrm{vir} R_\mathrm{vir} = \num{0.05}$, \num{0.11} and \num{0.10} respectively.
We note that though the halos are selected in the low-resolution DM-only simulation to have no major mergers between $z=2.5$ and $z=2$, they do have mergers in the high-resolution simulations with
mass ratio 1:5 at $z=2.3$ for halo A, %
mass ratio 1:5 at $z=2.2$ for halo B, %
and mass ratio 1:6 at $z=2.3$ for halo C. %
For each halo, we locate the central galaxy at $z=2$ in the reference zoom-in hydrodynamical simulation.
We then select all tracer particles (gas, stars and those accreted in supermassive black holes, see \citealt{cadiou_AccurateTracerParticles_2019}  for details) within the galaxy, defined by the region within $4 R_{1/2}$ at $z=2$.
We trace back the baryons to the initial conditions to find the baryonic Lagrangian patch of the galaxies\footnote{There are on average 8 tracers per cell in the initial conditions, and we consider any cell where more than 5 tracer particles end in the galaxy to belong to the patch.}. This baryonic Lagrangian patch is different from the DM patch one obtains when tracing back the DM particles instead \citep{liao_SegregationBaryonsDark_2017}.

\begin{figure*}
    \includegraphics[width=\textwidth]{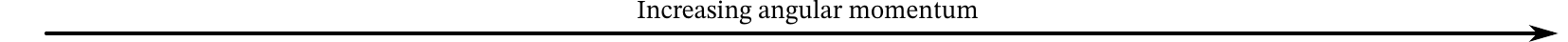}
    \includegraphics[width=\textwidth]{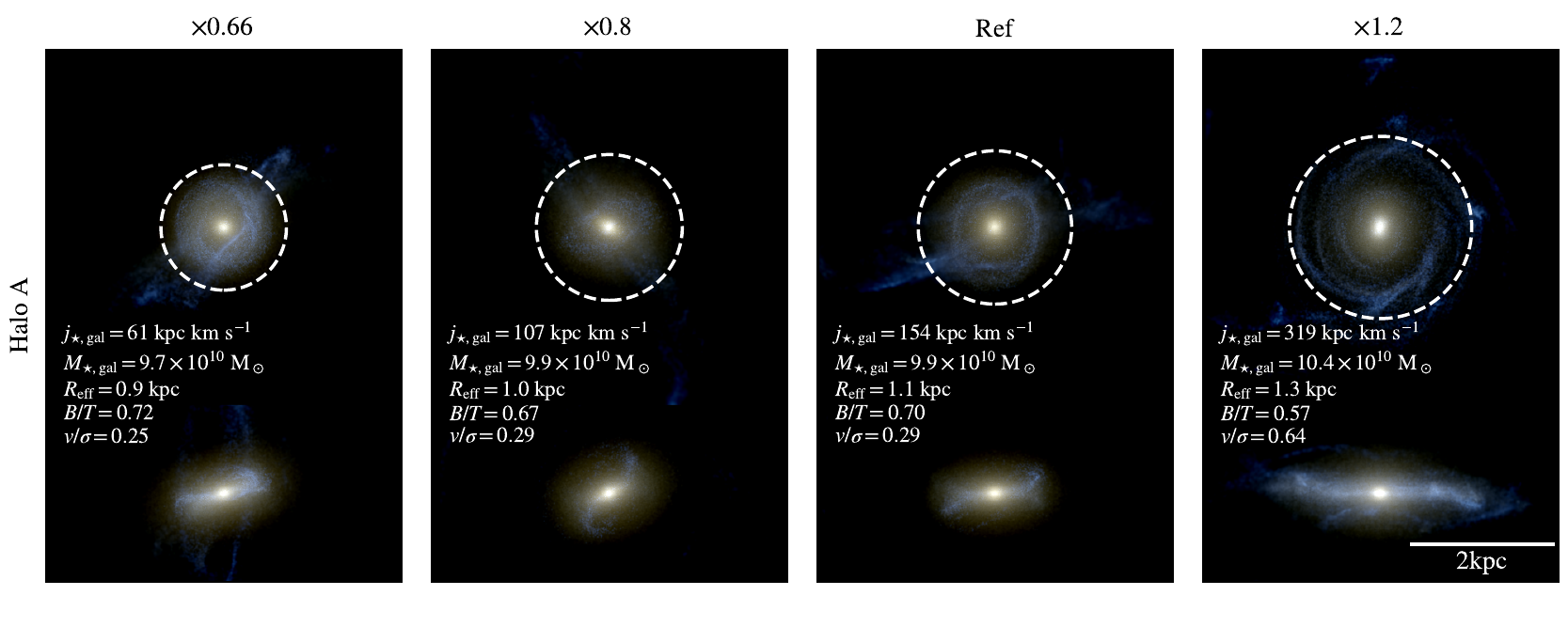}
    \caption{
        Mock images of the galaxy in halo A at $z=1.95$ sorted by increasing stellar angular momentum (from left to right).
        We show here the stellar specific angular momentum, mass, effective radius and bulge fraction total ratio measured \emph{in the galaxy}.
        The top row shows face-on mock images, the bottom row shows edge-on images.
        At fixed mass, galaxies with higher angular momentum have a more pronounced stellar disk, are more spatially extended, have a smaller bulge and a higher velocity-to-velocity dispersion $v/\sigma$ parameter.
        Conversely, galaxies with decreased stellar angular momentum are less disky, more compact and bulgy.
    }\label{fig:mock_images}
\end{figure*}
We scale the magnitude of the initial angular momentum with the genetic modification technique described above, applied to the baryonic patch at $z=200$. Due to the early-universe gravitational couplings between dark matter and baryons, this modification also changes the dark matter spin. For each of the three reference galaxies, we generate four additional galaxy initial conditions where the angular momentum $j_0$ of the patch has been scaled relative to the reference angular momentum $j_{0,\mathrm{ref}}$ by a factor $j_0/j_{0,\mathrm{ref}} = 0.66, 0.8, 1.2$ and $1.5$ respectively.
Such amplitudes are large enough to sample a wide range of stellar angular momentum, as we will show later, while still allowing precise control on the angular momentum in the evolved Universe within a few tens of percent \citep{cadiou_AngularMomentumEvolution_2021}.
We fix the $z=2$ halo mass when modifying the simulations by keeping the mean density of the entire dark matter halo Lagrangian patch fixed.

We use these genetically-modified initial conditions to perform 12 additional hydrodynamical zoom-in simulations. In total, we thus have a sample of 3 galaxies with 5 different angular momentum scenarios each, for a total of 15 simulations. However, we subsequently removed from the sample the special case of halo C$\times 1.5$ because the interpretation is hindered by a large delay in its assembly; this is discussed further in \cref{sec:results}.

At the end of the simulation, the host halos in the remaining 14 simulations have masses of $M_\mathrm{DM}\approx \SI{e12}{\Msun}$ and stellar mass within the virial radius of $ M_\star(<R_\mathrm{vir})\approx\SI{e11}{\Msun}$. We report in the left column of \cref{tab:masses} the halo masses for the three reference simulations, as well as their minimum and maximum value in the 11 modified simulations, showing that the values change from the reference value by 7\% at most once the single case of C$\times 1.5$ has been excluded. The middle column shows the stellar mass interior to the virial radius, which likewise is very stable, varying from the reference value by at most 10\%. The right column shows the stellar mass in the central galaxy, which is more variable (up to almost a factor 2). This is because the modification to the angular momentum changes the orbital trajectory of the satellites even after they have entered the virial radius, which in turn means they merge with the central galaxy at differing times relative to the end of the simulation.

\begin{table}
    \centering
    \caption{
        Minimum, maximum and reference values of the DM halo mass, the stellar mass within the virial radius, and the stellar mass in the central galaxy at the end of the simulations at $z=1.95$.
        Modifying the angular momentum keeps the halo mass and the stellar mass in the virial radius fixed within a few tens of percent.
        The mass of the central galaxy changes due to changes in the orbital trajectory of satellites, which advance or delay their merger.
    }\label{tab:masses}
    \begin{tabular}{llSSS}
        Halo &          & {$M_\mathrm{DM}$} & {$M_\star(<R_\mathrm{vir})$} & {$M_{\mathrm{gal},\star}$}  \\
             &          & \SI{e11}{\Msun}   & \SI{e10}{\Msun}              & \SI{e10}{\Msun}             \\
        \midrule                                                                                         
        A &       min &             9.8 &  9.8                           &            6.3              \\
            & reference &            10.4 & 10.9                           &            9.9               \\
            &       max &            10.4 & 11.2                           &           10.4               \\[.1cm]
        B&       min &            10.2 & 10.6                           &            4.5               \\
            & reference &            10.6 & 10.8                           &            8.7               \\
            &       max &            10.6 & 11.4                           &            8.7               \\[.1cm]
        C&      min &            10.0 & 09.9                           &            6.6               \\
            & reference &            10.8 & 11.2                           &            7.3               \\
            &       max &            10.9 & 11.2                           &            8.5               \\
    \end{tabular}
\end{table}

\section{Results}%
\label{sec:results}

Our goal is to quantify how changes in the initial angular momentum of the baryonic Lagrangian patch affect the angular momentum and observable galaxy properties at redshift $z=2$. For each of our simulations, we first measure the specific angular momentum of all the baryons, of all the stars and of all the DM within the virial radius at $z=2$, $j_\mathrm{b}$, $j_\star$ and $j_\mathrm{DM}$, including particles in satellite galaxies. The angular momentum is computed about the central galaxy using a shrinking sphere approach on the stellar density \citep{power_InnerStructureLCDM_2003}. In order to quantify how well we can control these angular momenta, from each modified simulation we calculate $j(z=2) / j_{\mathrm{ref}}(z=2)$ separately for stars, baryons and DM.
Figure~\ref{fig:AM_change__vs__stars_AM} compares this measured ratio to the imposed ratio in the initial conditions at $z=200$.
In all but one case (halo B$\times 0.66$), the change of stellar angular momentum at $z=2$ is in the direction expected (left panel): its value increases (resp.\ decreases) when we increase (resp.\ decrease) the initial angular momentum.
 For the one contrary case (halo B$\times 0.66$), we were unable to identify a single cause of the unexpectedly large angular momentum at $z=2$, but we note that this system has a complex merger history which may introduce non-linear interactions between the orbits of different sub-halos. We discuss this point further below.
 
 For halo A$\times 1.5$, the final angular momentum is increased but by a surprisingly large factor of $4$. In this case, the central galaxy is not relaxed and is undergoing a 1:2 merger event. At $z=2.0$, the companion is at its apocenter after reaching pericenter at $z = 2.06$. This causes the center of the central galaxy about which angular momentum is computed to be significantly offset from the center of mass of the system.
 Overall, based on this initial set of simulations, the relation between initial conditions and final stellar angular momentum is strong; the Spearman correlation coefficient between the initial modification and the measured change is \SI{83}{\percent}.

Our control of the angular momentum is somewhat weaker for baryons and DM, for which we measure a statistical correlation of $\approx\SI{70}{\percent}$ (see \cref{fig:AM_change__vs__stars_AM}, central and right panels). We have previously shown that particles in the outskirts of halos dominate the halo spin; therefore, small changes to the halo finder's boundary can cause apparently chaotic changes in the angular momentum of dark matter and gas \citep{cadiou_AngularMomentumEvolution_2021}. Stars, by contrast, typically occupy the deepest parts of the potential and their angular momentum is therefore less impacted by changes in the halo's boundary.

\begin{figure}
    \begin{subfigure}[b]{\columnwidth}
        \includegraphics[width=\columnwidth]{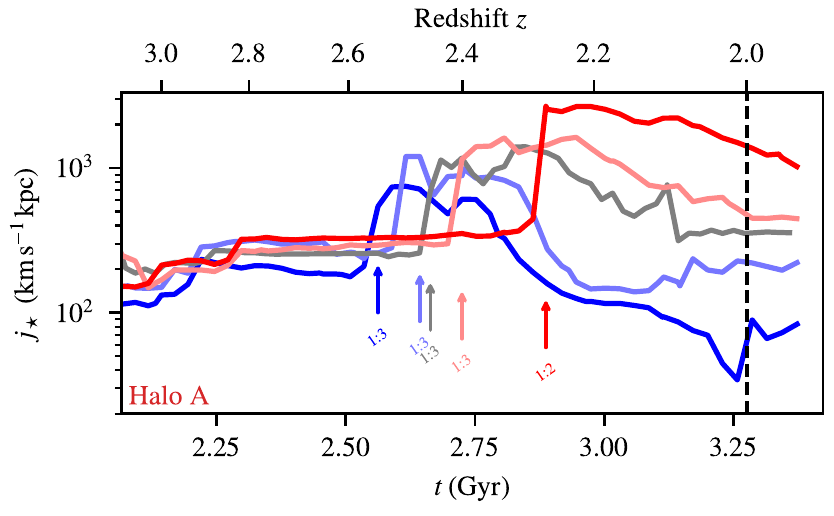}
    \end{subfigure}
    \begin{subfigure}[b]{\columnwidth}
        \includegraphics[width=\columnwidth]{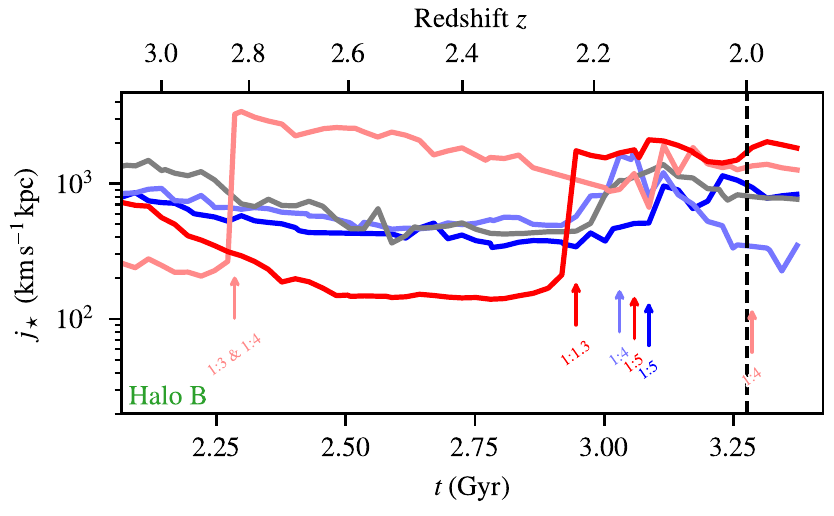}
    \end{subfigure}
    \begin{subfigure}[b]{\columnwidth}
        \includegraphics[width=\columnwidth]{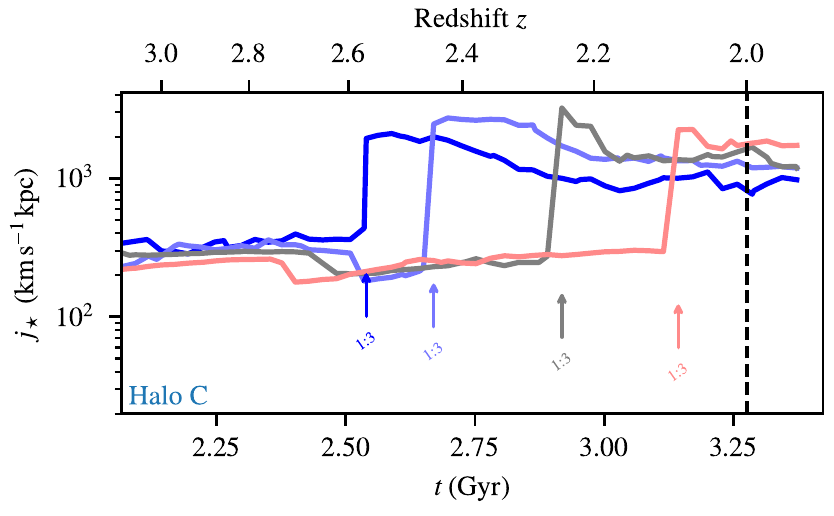}
    \end{subfigure}
    \caption{
        Evolution of the stellar angular momentum within the virial radius of halos A, B and C from top to bottom.
        We genetically modify the angular momentum in the initial conditions ($z=200$) of the region that contains all baryons that will comprise the galaxy at $z=2$.
        The initial angular momentum of the baryons is changed by a factor $0.66$, $0.8$, $1.2$ and $1.5$ (in dark blue, blue, red, dark red resp.) compared to a reference simulation (in grey).
        We show the time at which satellite galaxies cross the virial radius inward as coloured arrows together with their stellar mass ratio compared to the central galaxy.
        This increase (resp.\ decrease) of the \emph{initial} baryon angular momentum causes the specific angular momentum of the stars to increase (resp.\ decrease) by $z=2$. It also changes the infall time of satellites in a systematic way, due to changes in their orbital angular momentum. 
    }\label{fig:spAM__vs__time}%
    \label{fig:spAM__vs__time_A}%
    \label{fig:spAM__vs__time_B}%
    \label{fig:spAM__vs__time_C}%
\end{figure}

While the total angular momentum of stars inside the halo can be controlled, this quantity includes the effect of satellites. We now focus on a case where there are no major satellites, in order to investigate the angular momentum of the central galaxy itself. We count `major' satellites as having a stellar mass ratio more significant than 1:8; by this definition, galaxy A has no major satellite inside its virial radius at $z=2$.  This remains true in three of the four modifications (with $\times 0.66$, $\times 0.8$ and $\times 1.2$ angular momentum); in the case of $\times 1.5$, a 1:2 galaxy merger is delayed until after $z=2$ due to its increased orbital angular momentum.
In the cases where all major galaxy mergers have completed, the stellar mass and angular momentum within the virial radius is dominated by the central galaxy's disk.
We show in \cref{fig:mock_images} mock images in rest-frame $i,v,u$ bands for the red, green and blue channels respectively together with the galaxy's stellar specific angular momentum, stellar mass, stellar half-mass radius, bulge-to-total mass fraction and $v/\sigma$ parameter.
As we increase angular momentum in the initial conditions, the value of the galaxy's stellar angular momentum at $z=2$ increases while its stellar mass remains constant.
This  allows us to explore how the galaxy's angular momentum is reflected in its observable properties.

Qualitatively, the galaxy displays a ring-like structure in the lowest angular momentum simulation (left) that progressively disappears as angular momentum increases.
To understand its origin, we measure the angle $\cos\theta = \hat{\vec{j}}_\star \cdot \hat{\vec{j}}_\mathrm{orbit}$ between the direction of the central galaxy's angular momentum, $\hat{\vec{j}}_\star$, and the orbital angular momentum of the last major merger, $\hat{\vec{j}}_\mathrm{orbit} = \vec{r}\cross \vec{v} / |\vec{r} \cross \vec{v}|$. Here $\vec{r}$ and $\vec{v}$ are the position and bulk velocity of the satellite relative to the central galaxy. We average the angle along the satellite during its infall through the virial radius.
An angle of \SI{0}{\degree} corresponds to an orbit in the plane of the galaxy's disk, while an angle of \SI{90}{\degree} corresponds to an orbit perpendicular to it.
We find that, as we modify angular momentum upward, the orbit becomes increasingly coplanar: the mean angle during the infall through the virial radius (from \SI{60}{kpc} to \SI{20}{kpc}) is \num{77}, \num{71}, \num{63} and \SI{58}{\degree} for the $\times 0.66, 0.8$, reference and $\times 1.2$ cases respectively.
The mean angle measured just before the merger (at separation less than \SI{20}{kpc}) are
\num{52}, \num{50}, \num{31} and \SI{27}{\degree}.

We thus interpret the ring visible in \cref{fig:mock_images} (left panel) as the remnant of the last major merger that falls through the virial radius almost perpendicular (\SI{77}{\degree}) to the already-existing disk before realigning itself partially (\SI{52}{\degree}).
Conversely, in the increased angular momentum case (right panel), the orbit of the satellite upon passing through the virial radius is already more aligned with the galaxy's spin (\SI{58}{\degree}) and becomes almost coplanar by the time it merges (\SI{27}{\degree}).
As a consequence, modifications lead to a factor of $\times 6$ increase in stellar angular momentum from the lowest angular momentum scenario to the highest angular momentum one, while maintaining a fixed stellar mass. From lowest to highest angular momentum simulations, the half-mass radius $R_{1/2}$ increases by \SI{40}{\percent} from $0.9$ to $1.3\,\mathrm{kpc}$.
Similarly, the galaxy's bulge-to-total mass fraction decreases by \SI{25}{\percent}, while the post-merger $v/\sigma$ parameter increases by $\times 2.6$. While previous statistical studies have  suggested that morphology can be changed by the alignment of a merger
\citep{hopkins_HowDisksSurvive_2009,martin_RoleMergersDriving_2018,jackson_WhyExtremelyMassive_2020,zeng_FormationMassiveDisc_2021} our study provides the first direct test of this on individual galaxies in a cosmological environment.

The importance of satellites in delivering angular momentum, and the effect that angular momentum has on the orbital dynamics, can also be seen in \cref{fig:spAM__vs__time} where we plot the stellar angular momentum within the virial radius, $j_\star(r<R_\mathrm{vir})$, as a function of time. Halo A, the galaxy which we have just considered in detail, is shown in the upper panel. The large jumps in angular momentum correspond to the moment when massive satellite galaxies fall through the virial radius; we show infall times with arrows for all mergers with a stellar mass ratio larger than 1:8. (Note the galaxy mergers happen a few dynamical times later than the infall time, depending on the trajectory.)

The stellar angular momentum $j_\star(r<R_\mathrm{vir})$ jumps when satellite galaxies fall in, with the extent of the jump depending on both the trajectory of the orbit and its alignment relative to the existing stars.
This is particularly striking for halos A and C.
In these two cases, the timing of the infall changes substantially as we modify the initial angular momentum: increased (resp.\ decreased) angular momentum modifications delay (resp.\ hasten) the infall time by as much as \SI{500}{Myr}. In the case of halo C, the $\times 1.5$ case has been excluded because the infall time is so delayed that it does not occur by the end of the simulation, as we previously noted.

The case of halo B is more complicated, with satellite galaxies entering the virial radius until the end of the simulation at $z=2$.
In this case, the angular momentum modification also changes the order in which mergers take place. As we previously commented, this complexity leads to our sole example of an angular momentum modification which has an unexpected effect; the dark blue line represents the $\times 0.66$ case which should end with the lowest angular momentum but instead finishes with comparable angular momentum to the reference case. The complexity of the merger history precludes a full understanding, but we note that torques between satellites on either side of the halo boundary may have a role to play. Further investigation will require a larger sample of modified simulations.

In summary, our results show \emph{(i)} that angular momentum of stars in a given halo can be causally linked to the angular momentum of relevant patches of the initial conditions; \emph{(ii)} the angular momentum of the central galaxy also follows the expected trend, provided that there are no unmerged massive satellites; and \emph{(iii)} that the morphology of the central galaxy in the latter case changes significantly when the angular momentum is systematically altered.

\section{Discussion and conclusions}%
\label{sec:discussion}

In this paper, we tested for the first time whether the angular momentum of galaxies is predictably related to their initial conditions in the early universe.
We employed zoom-in numerical simulations with state-of-the-art physics to study the formation of three simulated galaxies.
We combined the angular momentum genetic modification technique \citep{cadiou_AngularMomentumEvolution_2021} with Lagrangian tracer particles \citep{cadiou_AccurateTracerParticles_2019} to reconstruct the Lagrangian region from which all baryons in a galaxy originate, which we then genetically modified to either increase or decrease the angular momentum they will acquire through torques with the cosmological environment.
Having re-simulated using the modified initial conditions, we measured the $z=2$ angular momentum within the virial radius of the halo and in the galaxy itself.

In this initial study, we chose to focus on three galaxies with masses $M_\star \approx \SI{e11}{\Msun}$ at $z=2$; at this epoch, such galaxies exhibit a range of morphologies in observations. We successfully controlled the angular momentum of the stars inside their virial radius in the evolved universe (\cref{fig:AM_change__vs__stars_AM}).
We conclude that the angular momentum of stars in the virial radius of individual galaxies can thus be predicted and controlled from the initial conditions, and that gravitational torques with the large-scale structure do play a significant role in determining the value of the final angular momentum. Our galaxy formation recipe includes state-of-the-art feedback prescriptions; thus we have established that, despite the importance of self-regulation \citep{ubler_WhyStellarFeedback_2014}, galaxies at our chosen mass and redshift retain memory of the angular momentum in their cosmological initial conditions.

In the case of a relaxed galaxy with no large satellites at $z=2$ (halo A), we were further able to control the central galaxy's angular momentum (\cref{fig:mock_images}). An increase of stellar angular momentum, $j_\star$, allows the formation of a more prominent and spatially extended disk. As stellar angular momentum increases, the $v/\sigma$ parameter also increases, while the bulge fraction $B/T$ decreases.
These trends are in agreement with observations at $z\approx 1.5$, which show an anticorrelation between the stellar angular momentum and bulge fraction at fixed stellar mass \citep{gillman_PeculiarMorphologiesHubbletype_2020} and a correlation with $v/\sigma$ parameter \citep{harrison_KMOSRedshiftOne_2017}.
For this galaxy, we also show that the prominence of the disk is regulated by its angular momentum content, in agreement with the fact that, at fixed stellar mass, spiral galaxies are more angular-momentum rich than elliptical ones \citep{1983IAUS..100..391F}.

Our work establishes a clear demonstration of cause-and-effect: the early universe controls the stellar angular momentum in a galaxy which, in turn, changes the morphology as parametrised by $B/T$ and $v/\sigma$.
In the future, we hope to increase the sample size to different stellar masses and observation redshifts to confirm that angular momentum is the key parameter driving scaling relations between $M_\star$, $j_\star$, $B/T$, $v/\sigma$, galaxy size and galaxy morphology.

By studying the stellar angular momentum time evolution, we showed in \cref{fig:spAM__vs__time} that, as satellites infall into the virial radius, the angular momentum undergoes significant changes. As a result, stellar angular momentum is sensitive to the infall of major satellites, the timing of which changes substantially as we modify the initial angular momentum. Our results support the idea that angular momentum jumps with mergers \citep{vitvitska_OriginAngularMomentum_2002,benson_GalaxyFormationSpanning_2010,benson_RandomwalkModelDark_2020a}, but additionally shows that the angular momentum in the initial conditions determines the impact parameter and timing of mergers in a deterministic way. This dependence of the merger history on initial angular momentum could be used to investigate how impact parameter and timing determine the effects of a merger \citep[e.g.][]{davies_GalaxyMergersCan_2022}.

During more complex periods of multiple mergers (as in the case of halo B), the interpretation of the origin of the stellar angular momentum is more ambiguous. Yet even in this case, we found that we can control the stellar angular momentum within the virial radius except when we tried to increase it by too large a factor ($\times 1.5$ in the initial conditions). This ability to reliably control angular momentum under challenging circumstances opens up the possibility of future simulations gaining insight into the internal physical processes of galaxy formation, for example by testing whether decreasing a galaxy's angular momentum drives the growth rate of the central supermassive black hole \citep[as suggested in][]{bellovary_RelativeRoleGalaxy_2013,curtis_ResolvingFlowsBlack_2016,angles-alcazar_GravitationalTorquedrivenBlack_2017}.

Our results also suggest that predicting stellar angular momentum and hence disk alignments \emph{ab initio} may be possible. We have shown that, once the patch of the initial conditions from which the baryons accrete is identified, the final stellar angular momentum closely tracks the patch's initial spin. Because the stars occupy the deepest part of the potential, their corresponding patch is likely to be predictable using approximate or machine learning methods \citep{lucie-smith_MachineLearningCosmological_2018,lucie-smith_InterpretableMachinelearningFramework_2019}. Current predictions based instead on the dark matter patch \citep[][Fig. 19]{porciani_TestingTidaltorqueTheory_2002,park_FormationMorphologyFirst_2022} suffer from particles in the outskirts dominating the angular momentum budget \citep{cadiou_AngularMomentumEvolution_2021}. Focussing  on baryonic patches could enable more precise predictions of the intrinsic alignment signal \citep[see][for a review]{kiessling_GalaxyAlignmentsTheory_2015}, lessening uncertainty in a key systematic for forthcoming weak lensing surveys.

\section*{Acknowledgements}

CC thanks C.~Pichon, Y.~Dubois and J.~Devriendt who provided useful feedback on the project. This project has received funding from the European Union's Horizon 2020 research and innovation programme under grant agreement No.~818085 GMGalaxies.
This work used computing equipment funded by DiRAC (\SI{10}{Mcpu.hr}, project dp160).
Data analysis was carried out on facilities supported by the Research Capital Investment Fund (RCIF) provided by UKRI and partially funded by the UCL Cosmoparticle Initiative.
Part of the project made use of the Infinity Cluster hosted by Institut d'Astrophysique de Paris. 
The work of HVP was partially supported by the Göran Gustafsson Foundation for Research in Natural Sciences and Medicine and the European Research Council (ERC) under the European Union's Horizon 2020 research and innovation programme (grant agreement no. 101018897 CosmicExplorer).
This work has been enabled by support from the research project grant `Understanding the Dynamic Universe' funded by the Knut and Alice Wallenberg Foundation under Dnr KAW 2018.0067.
AP was supported by the Royal Society.
The analysis was carried out using
\textsc{colossus} \citep{diemer_COLOSSUSPythonToolkit_2018},
\textsc{jupyter} notebooks \citep{kluyver_JupyterNotebooksPublishing_2016},
\genetic{} 1.3.5 \citep{pontzen_PynbodyGenetICVersion_2022},
\textsc{matplotlib} \citep{hunter_Matplotlib2DGraphics_2007},
\textsc{numpy} \citep{harris_ArrayProgrammingNumPy_2020},
\textsc{pynbody} \citep{pontzen_PynbodyNBodySPH_2013},
\textsc{python},
\textsc{tangos} \citep{pontzen_TANGOSAgileNumerical_2018} and
\textsc{yt} \citep{turk_YtMulticodeAnalysis_2011}.

\section*{Author contributions}

The main roles of the authors were, using the CRediT (Contribution Roles Taxonomy) system (\url{https://authorservices.wiley.com/author-resources/Journal-Authors/open-access/credit.html}):

{\bf CC:} conceptualisation; methodology; validation; investigation; data curation; formal analysis; writing -- original draft; visualisation.

{\bf AP:} conceptualisation; methodology; funding acquisition; writing~-- review and editing; validation.

{\bf HVP:} conceptualisation; writing -- review and editing; validation.

\section*{Data availability}
The data underlying this article will be shared on reasonable request to the corresponding author.

\bibliographystyle{mnras}
\bibliography{authors}

\appendix

\section{Technical description of the simulation setup}
\label{sec:technical_details}
The simulations are started with a coarse grid of $512^3$ (level 7) and several nested grids with increasing levels of refinement up to level 12, corresponding to a DM mass resolution of respectively \SI{8.2e8}{\Msun}, \SI{1e8}{\Msun}, \SI{1.3e7}{\Msun} and \SI{1.6e6}{\Msun}.

Particles (DM, stars, black holes) are moved with a leap-frog scheme, and to compute their contribution to the gravitational potential, their mass is projected onto the mesh with a cloud-in-cell interpolation.
Gravitational acceleration is obtained by computing the gravitational potential through the Poisson equation numerically obtained with a conjugate gradient solver on levels above 12, and a multigrid scheme~\citep{guillet_SimpleMultigridScheme_2011} otherwise.
We include hydrodynamics in the simulations, which system of non-linear conservation laws is solved with the MUSCL-Hancock scheme~\citep{vanleer_UltimateConservativeDifference_1979} using a linear reconstruction of the conservative variables at cell interfaces with minmod total variation diminishing scheme, and with the use of the HLLC approximate Riemann solver~\citep{toro_RestorationContactSurface_1994} to predict the upstream Godunov flux.
We allow the mesh to be refined according to a quasi-Lagrangian criterion: if $\rho_{\rm DM}+\rho_{\rm b}/f_{\rm b/DM}>8 m_{\rm DM,res}/\Delta x^3$, where $\rho_{\rm DM}$, and $\rho_{\rm b}$ are respectively the DM and baryon density (including stars, gas, and supermassive black holes (SMBHs)), and where $f_{\rm b/DM}$ is the universal baryon-to-DM mass ratio.
Conversely, an oct (8 cells) is de-refined when this local criterion is not fulfilled.
The maximum level of refinement is also enforced up to 4 minimum cell size distance around all SMBHs.
The simulations have a roughly constant proper resolution of \SI{35}{pc} (one additional maximum level of refinement at expansion factor $0.1$ and $0.2$ corresponding to a maximum level of refinement of respectively 19 and 20), a star particle mass resolution of $m_{\star,\rm res}=\SI{1.1e4}{\Msun}$, and a gas mass resolution of \SI{2.2e5}{\Msun} in the refined region.
Each Monte-Carlo tracer samples a mass of $m_\mathrm{t} = \SI{3.9e4}{\Msun}$ ($N_\mathrm{tot}\approx\num{1.1e7}$ particles).
There is on average 0.3 tracer per star and 22 per initial gas resolution element.
Cells of size \SI{35}{pc} and gas density of \SI{40}{cm^{-3}} contain on average one tracer per cell.

The simulations include a metal-dependent tabulated gas cooling function following~\cite{sutherland_CoolingFunctionsLowDensity_1993} for gas with temperature above $T> \SI{e4}{K}$.
The metallicity of the gas in the simulation is initialised to $Z_0 = \SI{e-3}{Z_\odot}$ to allow further cooling below \SI{e4}{K} down to $T_{\min} = \SI{10}{K}$ \citep{rosen_GlobalModelsInterstellar_1995}.
Reionisation occurs at $z=8.5$ using the~\cite{haardt_RadiativeTransferClumpy_1996} UV background model and assuming gas self-shielding above \SI{e-2}{cm^{-3}}.
Star formation is allowed above a gas number density of $n_0=\SI{10}{cm^{-3}}$ with a Schmidt law, and with an efficiency $\epsilon_{\rm ff}$ that depends on the gravo-turbulent properties of the gas \citep[for a comparison with a constant efficiency see][]{nunez-castineyra_CosmologicalSimulationsSame_2021}.
The stellar population is sampled with a~\cite{kroupa_VariationInitialMass_2001} initial mass function.
The mass loss fraction from supernovae explosions is $\eta_{\rm SN}=\SI{32}{\percent}$ with a metal yield (mass ratio of the newly formed metals over the total ejecta) of $0.05$.
Type II supernovae are modelled with the mechanical feedback model of~\cite{kimm_SimulatingStarFormation_2015} with a boost in momentum due to early UV pre-heating of the gas following~\cite{geen_DetailedStudyFeedback_2015}.
The simulations also track the formation of SMBHs and their energy release through AGN feedback.
SMBH accretion assumes an Eddington-limited Bondi-Hoyle-Littleton accretion rate in jet mode (radio mode) and thermal mode (quasar mode) using the model of~\cite{dubois_FeedingCompactBulges_2012}.
The jet is modelled self-consistently by following the \AM{} of the accreted material and the spin of the black hole \citep{dubois_BlackHoleEvolution_2014}.
The radiative efficiency and spin-up rate of the SMBH is computed assuming the radiatively efficient thin accretion disk from~\cite{shakura_BlackHolesBinary_1973} for the quasar mode, while the feedback efficiency and spin-up rate in the radio mode follows the prediction of the magnetically choked accretion flow model for accretion disks from~\cite{mckinney_GeneralRelativisticMagnetohydrodynamic_2012}.
SMBHs are created with a seed mass of \SI{e5}{\Msun}.
For the exact details of the spin-dependent SMBH accretion and AGN feedback, see~\cite{dubois_IntroducingNEWHORIZONSimulation_2021}.

\bsp{}%
\label{lastpage}
\end{document}